\documentclass[aps,pra,twocolumn,amsmath,amssymb]{revtex4-2}

\usepackage{graphicx}
\usepackage{amsmath}
\usepackage[colorlinks=false]{hyperref}
\usepackage{dsfont}
\usepackage{dirtytalk}

\begin{document}

\title{Squeezing by Critical Speeding-up: Applications in Quantum Metrology}
\author{Karol Gietka}
\email{karol.gietka@oist.jp}
\address{Quantum Systems Unit, Okinawa Institute of Science and Technology Graduate University, Onna, Okinawa 904-0495, Japan}


\begin{abstract}
We present an alternative protocol allowing for the preparation of critical states that instead of suffering from the critical slowing-down benefits from the critical speeding-up. Paradoxically, we prepare these states by going away from the critical point which allows for the speed-up. We apply the protocol to the paradigmatic quantum Rabi model and its classical oscillator limit as well as the Lipkin-Meshkov-Glick model. Subsequently, we discuss the application of the adiabatic speed-up protocol in quantum metrology and compare its performance with critical quantum metrology. We show that critical quantum metrology with the Lipkin-Meshkov-Glick model cannot even overcome the standard quantum limit, and we argue that, even though critical metrology protocols can overcome it in some cases, critical metrology is a suboptimal metrological strategy. Finally, we conclude that systems exhibiting a phase transition are indeed interesting from the viewpoint of quantum technologies, however, it may not be the critical point that should attract the most attention.
\end{abstract}
\maketitle


\section{Introduction}
The recent experimental progress in isolating and manipulating quantum systems has brought us on the verge of entering the era of quantum technologies. One of its key aspects will be the reliable and robust preparation of quantum states. Especially, high-fidelity preparation of critical ground states---ground states close to a critical point of a quantum phase transition~\cite{2005_Quantumcriticality}---has been identified as a key ingredient in many quantum technologies such as quantum metrology~\cite{2008_Zanardi_CriticalityasresourceinQM,2016HaukeZoller,2018_QuantumCriticalMetrology,2021HaukeEnt,2021dicandia2021critical} and quantum heat engines~\cite{2016_Fazio_criticalheatengine,2020Mossyengine}. The reason behind the universality of these states is mainly the high level of nonclassical correlations. These correlations may come in the form of squeezing, spin squeezing, and entanglement. Therefore the critical states might be also used to study fundamental aspects of quantum mechanics. Unfortunately, the critical ground states are also difficult to prepare. This is typically caused by the closing of the energy gap between the ground state and the excited states near a critical point. If an adiabatic process---a process during which the instantaneous state is the eigen state of the system---is used to create the critical ground state, its duration will be very long. This is a consequence of the adiabatic theorem which states that in order to remain in the instantaneous ground state, the rate at which the Hamiltonian is being changed has to be much smaller than the instantaneous energy gap. Therefore if the energy gap decreases, the rate at which the Hamiltonian changes has to decrease as well. Moreover, close to the critical point, instability of the adiabatic ramp might push the system beyond the critical point and a phase transition will occur. To circumvent the critical slowing-down, one can resort to the shortcuts to adiabaticity~\cite{2013_ShortcutsToAdiabaticity,2019_ShortcustToAdiabaticity}. This approach involves a number of techniques. One of the most effective is the counterdiabatic driving~\cite{2013_CounterdiabaticDriving} which relies on adding an extra term to the Hamiltonian of the system and canceling the contributions from the excited states. As a consequence, the instantaneous state remains the ground state throughout the entire process. In principle, the counter-diabatic driving might be used to prepare critical ground states in arbitrarily short times; however, the energy cost of such a protocol might be very high~\cite{2017_Costofshortcuts}. This can be intuitively understood as a manifestation of the time-energy uncertainty relation. Alternative protocols rely on the so-called bang-bang protocols~\cite{2020_bang}. This approach is based on sudden changes of Hamiltonian parameters and often exploits squeezing following a nonequilibrium phase transition. Despite their simplicity, they allow for a high-fidelity ground state preparation in relatively short times. The drawback of bang-bang techniques is the requirement to abruptly change parameters and very precise moments of (some) bangs. Sudden quenches may often destroy the physical systems---for instance, due to the heating as in the atom-cavity experiments---while the inability to apply the bangs at the right moment will deteriorate quickly the fidelity as the evolution following a sudden quench is typically very fast. 

In this work, we present an alternative method for the preparation of critical ground states that relies on adiabatic driving the system away from the critical point. Such a technique allows us to open the energy gap as we approach the target state which results in the critical speeding-up. In fact, this method is based on the preparation of rotated---rotation in a phase space--critical ground states which can be then transformed into the true critical ground states through a simple $\pi/2$ rotation in the phase space picture. The speed-up protocol is a combination of an adiabatic protocol and (optionally) a bang-bang protocol but eliminates some of their drawbacks. We show how it can be used in the paradigmatic quantum Rabi model and the Lipkin-Meshkov-Glick model. Subsequently, we discuss the application of the adiabatic speed-up protocol in quantum metrology, we compare it with critical quantum metrology, and argue that critical metrology is a suboptimal metrological strategy as it exploits an ineffective way of creating nonclassical correlations. In particular, we show that critical metrology with the Lipkin-Meshkov-Glick model cannot even overcome the standard quantum limit. Finally, we discuss the possibility of harnessing the speed-up protocol in other systems exhibiting quantum phase transitions and using it in other quantum technologies.


\section{Adiabatic Critical Speeding-up}
In this section, we explain the operating mechanism of the critical speeding-up. The starting point is the classical oscillator limit of the quantum Rabi model. Subsequently, we discuss the general quantum Rabi model, and finally the Lipkin-Meshkov-Glick model.


\subsection{Quantum Rabi Model}
Quantum Rabi model is a paradigmatic model describing a harmonic oscillator with frequency $\omega$ coupled to a spin-$1/2$ with frequency $\Omega$ (we set $\hbar =1$ throughout the entire manuscript)
\begin{align}\label{eq:QRM}
  \hat H_{\mathrm{QRM}} = \omega \hat a^\dagger \hat a + \frac{\Omega}{2}\hat\sigma_z+\frac{g}{2}(\hat a + \hat a^\dagger)\hat\sigma_x,
\end{align}
where $g$ is the coupling constant. Here we have introduced creation and annihilation operators of the harmonic oscillator, $\hat a^\dagger$ and $\hat a$, and Pauli matrices $\hat \sigma_i$. The quantum Rabi model exhibits a (superradiant) phase transition if $\Omega>\omega$ from a single minimum phase ($g<g_c\equiv\sqrt{\omega\Omega}$) to a double minimum phase ($g>g_c\equiv\sqrt{\omega\Omega}$), where $g_c$ is the critical coupling strength. In the equivalent of the thermodynamic limit $\omega/\Omega\rightarrow 0$ (the classical oscillator limit~\cite{2012_FeshkeDickemodel,2015_QPT_Rabi}) the energy gap closes at the critical point and the Hamiltonian can be described by
\begin{align}\label{eq:qrm_lim}
  \hat H_{\mathrm{QRM}} = \omega \hat a^\dagger \hat a + \frac{\Omega}{2}\hat\sigma_z+\frac{g^2}{4\Omega}(\hat a + \hat a^\dagger)^2\hat\sigma_z.
\end{align}
If $\langle\hat \sigma_z \rangle < -g_c^2/g^2$, the above Hamiltonian does not have an equilbrium state in the critical phase. This becomes apparent when we rewrite it using the position and momentum operators, $\hat X = (\hat a +\hat a^\dagger)/\sqrt{2}$ and $\hat P = (\hat a - \hat a^\dagger)/\sqrt{-2}$,
\begin{align}\label{eq:qrm_col}
  \hat H_{\mathrm{QRM}} = \frac{\omega}{2} \hat P^2 + \frac{\Omega}{2}\hat\sigma_z+\frac{\omega}{2}\left(1+\frac{g^2}{g_c^2}\hat\sigma_z\right)\hat X^2.
\end{align}
The lack of the equilibrium state is often discussed as the limitation of the above approach; however, in the limit of $\omega/\Omega\rightarrow 0$ the ground state energy in the double minimum phase tends to $-\infty$ and the Hamiltonian describes an inverted harmonic oscillator~\cite{2021_GietkaIHO}. In order to find the ground state for $\omega/\Omega \ll 1$, the Hamiltonian is typically displaced to one of the minima in the critical phase~\cite{2015_QPT_Rabi} and the effective Hamiltonian resembles that from equation~\eqref{eq:qrm_col}. In this work, however, we focus only on the single minimum phase. 

\begin{figure*}[htb!]
 \centering
\includegraphics[width=0.8\textwidth]{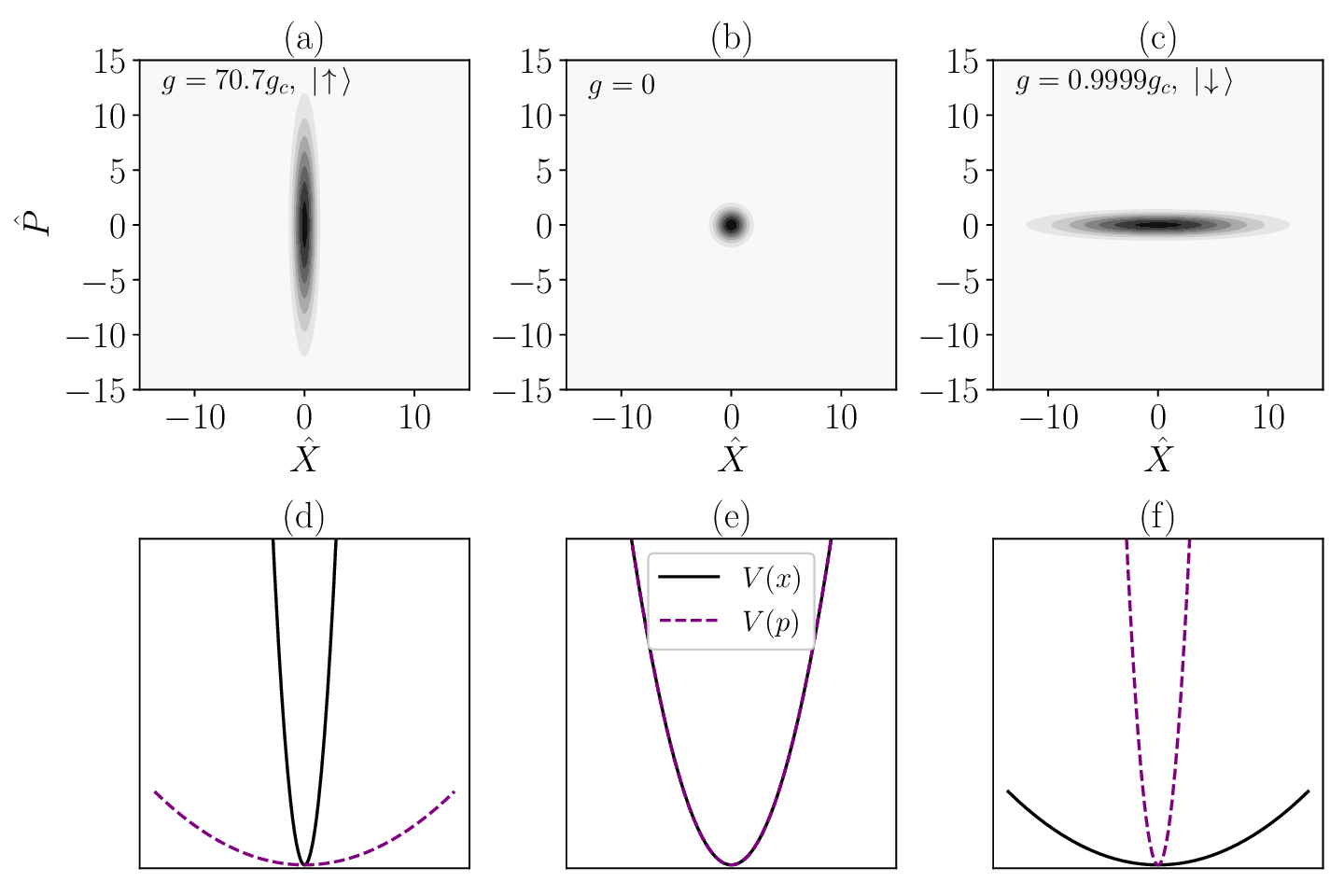}
\caption{In the quantum Rabi model, the same amount of squeezing (visible in the Husimi function~\cite{1940Husimi}) of a classical state (b) can be achieved by going to the critical state (c) or going away from the critical state (a). In the latter case, the initial spin has to flip before increasing the coupling strength. (d), (e), and (f), depicts effective position (solid black line) and momentum (dashed red line) potentials associated with states from (a), (b), and (c), respectively. (a) and (c) [(d) and (f)] are related by a simple rotation in the phase space (Fourier transform) whose generator is $\hat a^\dagger \hat a$.}
\label{fig:fig1}
\end{figure*}

The ground state of Hamiltonian \eqref{eq:qrm_col} can be easily found to be
\begin{align}
  |\mathrm{GS}\rangle_{\mathrm{QRM}} = \hat S(\xi) |0 \rangle \otimes |\!\downarrow \,\rangle,
\end{align}
where $\hat S(\xi) \equiv \exp\{(\xi/2)(\hat a^\dagger)^2-(\xi^*/2)\hat a^2\}$ is the squeeze operator with $\xi = -\frac{1}{4} \ln\{1-(g/g_c)^2\}$ being the squeezing parameter which is real only for $g < g_c$. In the above equation $|0\rangle$ is the vacuum state of the harmonic oscillator and $|\!\downarrow\,\rangle$ is the spin-down state (the ground state of $\hat \sigma_z$). The energy gap can be easily read from equation~\eqref{eq:qrm_col} and is equal to $\omega\sqrt{1-g^2/g_c^2}$ which vanishes at $g=g_c$. This can be easily understood because for $g=g_c$ the Hamiltonian describes a particle moving in a free space. The free particle eigen states are eigen states of momentum operator which are extremely nonclassical states. They saturate the Heisenberg position-momentum relation $\Delta \hat X \Delta \hat P = 1/2$ but are extremely squeezed $\Delta \hat P /\Delta \hat X \rightarrow 0 $. Once we understand that, we open the highway for the adiabatic critical ground state preparation. Since the critical ground states are simply squeezed, if we were able to increase the frequency of the harmonic oscillator from equation~\eqref{eq:qrm_col}, we could prepare a squeezed state in \emph{position} instead of \emph{momentum}. In this case, the energy gap would open which would allow for critical speeding-up of the adiabatic process. Transforming a state squeezed in position to a state squeezed in momentum would amount to setting $g=0$ and waiting for time $\omega t = \pi/2$ and performing a sudden quench (bang) to a desired $g$ for which the rotated state is the ground state. The intuitive depiction of this mechanism is presented in figure~\ref{fig:fig1}.

\begin{figure}[htb!]
 \centering
\includegraphics[width=0.4\textwidth]{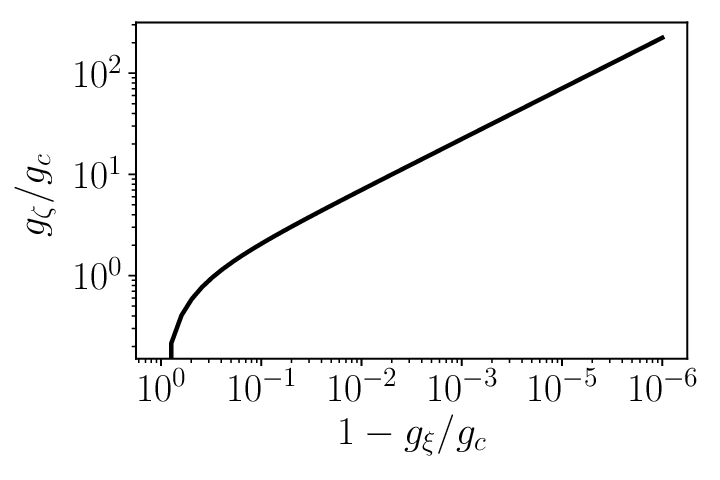}
\caption{Condition from equation~\eqref{eq:speedupcond}. The same amount of squeezing can be created by approaching the critical point $g_c$ or departing from it.}
\label{fig:fig2}
\end{figure}

Unfortunately, the coupling parameter enters Hamiltonian \eqref{eq:qrm_col} in the second power which suggests that in order to go away from the critical point one would have to make the coupling strength imaginary. However if the initial state is not the spin-down state but the spin-up state---so the excited state of the spin---the Hamiltonian becomes
\begin{align}\label{eq:qrm_col_up}
  \hat H_{\mathrm{QRM}}^\uparrow = \frac{\omega}{2} \hat P^2 + \frac{\omega}{2}\left(1+\frac{g^2}{g_c^2}\right)\hat X^2
\end{align}
instead of
\begin{align}\label{eq:qrm_col_down}
  \hat H_{\mathrm{QRM}}^\downarrow = \frac{\omega}{2} \hat P^2 + \frac{\omega}{2}\left(1-\frac{g^2}{g_c^2}\right)\hat X^2.
\end{align}
This means that simply by flipping the spin it is possible to increase the frequency of the effective harmonic oscillator instead of decreasing it. Since flipping a spin is a standard and well-understood tool in quantum technologies, critical speeding up could be achieved by a tiny modification of the standard protocols involving the quantum Rabi model. The state whose spin is excited but the part describing the harmonic oscillator is in its ground state is
\begin{align}
  |\mathrm{ES}\rangle_{\mathrm{QRM}} = \hat S(\zeta) |0 \rangle \otimes |\!\uparrow \,\rangle,
\end{align}
where $\hat S(\zeta) \equiv \exp\{(\zeta/2)(\hat a^\dagger)^2-(\zeta^*/2)\hat a^2\}$ is the squeeze operator with $\zeta = -\frac{1}{4} \ln\{1+(g/g_c)^2\}$ being the squeezing parameter which is always real. The energy gap can be easily read from equation~\eqref{eq:qrm_col_up} and is equal to $\omega\sqrt{1+g^2/g_c^2}$. Therefore, in order to obtain an equally squeezed state by closing or opening the energy gap, the squeezing parameter $\xi$ has to be equal to inverse of the squeezing parameter $\zeta$ which leads to the following condition 
\begin{align}\label{eq:speedupcond}
 \xi = \frac{1}{\zeta} \rightarrow \frac{g_\zeta}{g_c} = \sqrt{\frac{g_\xi^2}{g_c^2-g_\xi^2}},
\end{align}
where $g_\xi\leq g_c$ and $g_\zeta$ is the coupling strength for the spin-down and spin-up case, respectively. Here we can identify a possible drawback of the speed-up protocol. The price paid for opening the energy gap might be the necessity to increase the coupling far beyond the critical value. The condition from the equation~\eqref{eq:speedupcond} is presented in figure~\ref{fig:fig2}.

\begin{figure*}[htb!]
 \centering
\includegraphics[width=0.8\textwidth]{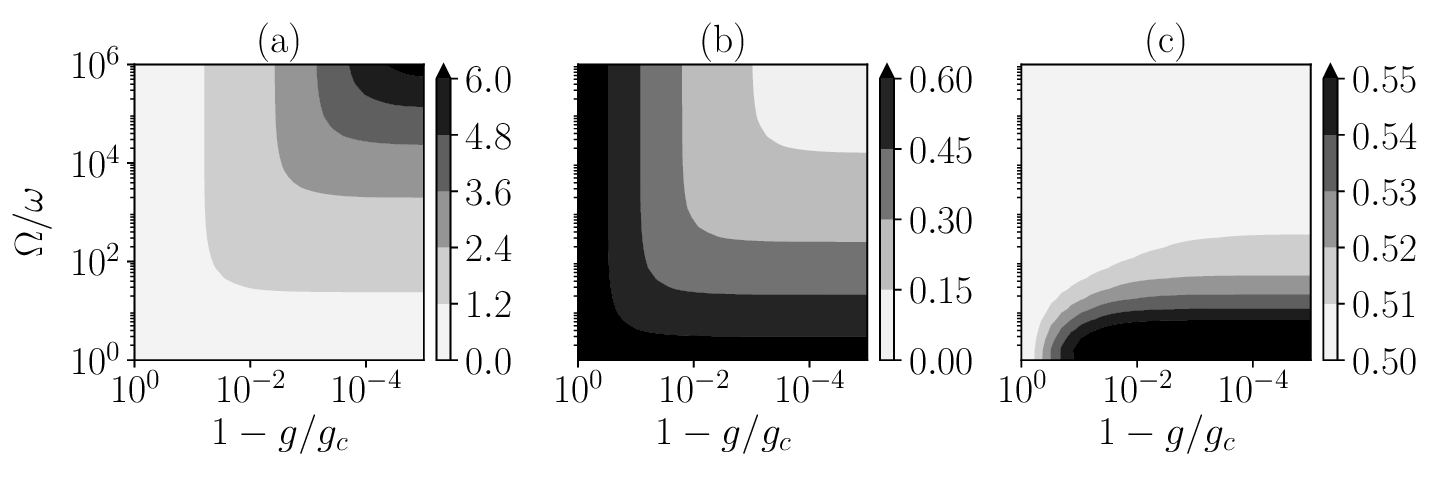}
\caption{Characterizing the ground state of the Rabi model as a function of $\Omega/\omega$ and $1-g/g_c$. (a), (b), and (c) shows $\Delta X$, $\Delta P$, and $\Delta X \Delta P$, respectively. The ground state is almost always a perfectly squeezed vacuum, i.e., $\Delta X \Delta P=1/2$. }
\label{fig:fig3}
\end{figure*}

So far we have discussed the classical oscillator limit of the quantum Rabi model. However, we have to consider the quantum Rabi model in a general case. This is a slightly more difficult task because we do not have an analytical (and suitable) form of the eigen states. Nevertheless, by performing numerical calculations and observing how the results converge to the analytical results in the classical oscillator limit, we can build up a physical intuition and phenomenologically understand what is happening. In order to do that, we will calculate standard deviations
\begin{align}
  \Delta O \equiv \sqrt{\langle \hat O^2\rangle - \langle \hat O \rangle^2},
\end{align}
of $\hat X$ and $\hat P$ operators as a function of $g/g_c$ and $\omega/\Omega$ for the ground state of the system as well as a state for which the spin is in its excited state and the harmonic oscillator is in its ground state. The results of the numerical simulations (exact diagonalization) for the ground state are presented in figure~\ref{fig:fig3}.

The results from figure~\ref{fig:fig3} mean that for every $\Omega/\omega$ there is a limit of how much the state can be squeezed. This suggests that the classical oscillator limit can be applied to the quantum Rabi model under extra conditions that relate $\Omega/\omega$ and $g/g_c$ . In order to find these conditions, we will apply the Schrieffer-Wolff transformation defined as
\begin{align}
  \hat U_{\mathrm{SW}} = \exp\big\{i \frac{g}{g_c} \frac{\sqrt{\omega}}{\sqrt{\Omega}} \left(\hat a^\dagger + \hat a\right)\frac{\hat \sigma_y}{2}\big\}.
\end{align}
This transformation rotates the spin around the $y$ axis by an angle $\frac{g}{g_c}\frac{\sqrt{\omega}}{\sqrt{\Omega}}(\hat a^\dagger + \hat a)$, and displaces the state of the harmonic oscillator by $i\frac{g}{g_c}\frac{\sqrt{\omega}}{\sqrt{\Omega}}\frac{\hat \sigma_y}{2}$. By applying this transformation to the quantum Rabi Hamiltonian~\eqref{eq:QRM}, we obtain
\begin{widetext}
\begin{align}
\begin{split}
  \hat H_{\mathrm{QRM}} =&\,\, \omega \left(\hat a^\dagger +i \frac{g}{g_c}\frac{\sqrt{\omega}}{\sqrt{\Omega}} \frac{\hat \sigma_y}{2}\right) \left(\hat a -i\frac{g}{g_c}\frac{\sqrt{\omega}}{\sqrt{\Omega}} \frac{\hat \sigma_y}{2}\right) + \frac{\Omega}{2}\left(\cos\left(\frac{g}{g_c}\frac{\sqrt{\omega}}{\sqrt{\Omega}}(\hat a^\dagger + \hat a)\right)\hat \sigma_z - \sin\left(\frac{g}{g_c}\frac{\sqrt{\omega}}{\sqrt{\Omega}}(\hat a^\dagger + \hat a)\right)\hat \sigma_x\right)\\
  &+ \frac{g}{2}\left(\hat a^\dagger+ \hat a \right)\left(\cos\left(\frac{g}{g_c}\frac{\sqrt{\omega}}{\sqrt{\Omega}}(\hat a^\dagger + \hat a)\right)\hat \sigma_x + \sin\left(\frac{g}{g_c}\frac{\sqrt{\omega}}{\sqrt{\Omega}}(\hat a^\dagger + \hat a)\right)\hat \sigma_z\right).
  \end{split}
\end{align}
\end{widetext}
If $\frac{g}{g_c}\frac{{\omega}}{{\Omega}}\ll1$ the above expression can be expanded in the Taylor series (the trigonometric functions have to be expanded to the second order thus $\frac{g}{g_c}\frac{\sqrt{\omega}}{\sqrt{\Omega}}\ll1$ is too stringent) to give
\begin{align}
  \hat H_{\mathrm{QRM}} =\,\, \omega \hat a^\dagger \hat a + \frac{\Omega}{2}\hat \sigma_z + \frac{\omega}{4}\frac{g^2}{g_c^2}(\hat a^\dagger + \hat a)^2\hat \sigma_z,
\end{align}
which is the classical oscillator limit from equation~\eqref{eq:qrm_lim}. This suggests that whenever $\frac{g}{g_c}\frac{{\omega}}{{\Omega}}\ll1$, we can use the classical oscillator limit Hamiltonian to easily find the spectrum. 

\begin{figure*}[htb!]
 \centering
\includegraphics[width=0.8\textwidth]{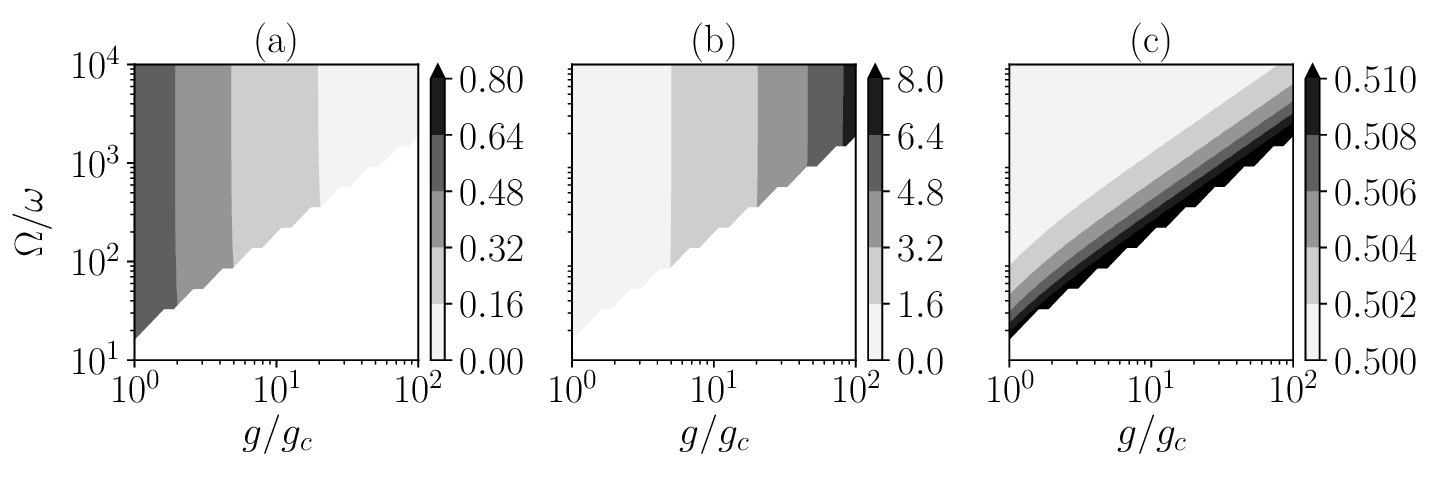}
\caption{Characterizing the eigen state of the quantum Rabi model whose spin is excited but the harmonic oscillator is in the ground state as a function of $\Omega/\omega$ and $g/g_c$ under the condition of $\frac{g}{g_c}\frac{{\omega}}{{\Omega}} < 0.05$. (a), (b), and (c) shows $\Delta X$, $\Delta P$, and $\Delta X \Delta P$, respectively. The state is always a perfectly squeezed vacuum if $\frac{g}{g_c}\frac{{\omega}}{{\Omega}} \ll 1$. The white color blocks are the regions where the spin-up sector starts to mix with the spin-down sector as can be seen in (c). See figure~\ref{fig:fig3} for comparison.}
\label{fig:fig5}
\end{figure*}

According to the calculation above, we can also use the classical oscillator limit of the quantum Rabi model to find the state which is an excited state of the spin and the ground state of the harmonic oscillator once $\frac{g}{g_c}\frac{{\omega}}{{\Omega}}\ll1$. This time, however, not satisfying this condition by increasing further $g$ means that the energies of the spin-up sector can no longer be separated from the spin-down sector. This will result in a possibility of transition from the spin-up to the spin-down state and the squeezing will be perturbed (see section~\ref{sec:QRMadiabatic} for more details). Therefore, we now find the excited state of the spin and the ground state of the harmonic oscillator as a function of $g/g_c$ and $\omega/\Omega$, but only for $\frac{g}{g_c}\frac{{\omega}}{{\Omega}} < 0.05$ (as we want to see the breaking of the approximation). The results of the numerical simulations (exact diagonalization) are presented in figure~\ref{fig:fig5}. These results resemble very much the results presented in figure~\ref{fig:fig3} but $\Delta X$ behaves as $\Delta P$ [\ref{fig:fig3}(a) and \ref{fig:fig5}(b)] and $\Delta P$ behaves as $\Delta X$ [\ref{fig:fig3}(b) and \ref{fig:fig5}(a)], as expected from the classical oscillator limit.

So far we have shown that it is possible to create the (rotated) ground state of the quantum Rabi model by going away from the critical point. Now, we will show that the same is also possible in the context of the Lipkin-Meshkov-Glick model


\subsection{Lipkin-Meshkov-Glick Model}\label{sec:LMG}
Lipkin-Meshkov-Glick model~\cite{LIPKIN1965188} is a paradigmatic model originally proposed to describe shape phase transitions in atomic nuclei. This model can be also used to describe two-mode systems of interacting bosons such as spinor condensates or two-site Bose-Hubbard model, and turns out to be the fast-oscillator limit of the Dicke model~\cite{2012_FeshkeDickemodel}. The Lipkin-Meshkov-Glick Hamiltonian can be expressed as
\begin{align}\label{eq:LMG}
  \hat H_{\mathrm{LMG}} = \omega \hat S_z - \frac{g}{N}\hat S_x^2,
\end{align}
where $\hat S_i = \sum_{i=1}^N \hat \sigma_i/2$ are the collective spin operators. Here $\omega$ is the energy separation between the levels of a single spin and $g$ is the interaction strength. The Lipkin-Meshkov-Glick model exhibits a phase transition from a single minimum phase to a double minimum phase for $g_c=\omega$. This can be easily seen if we apply the Holstein-Primakoff transformation. Under this transformation the collective spin operators can be expressed in terms of a single bosonic mode
\begin{align}
\begin{split}
  \hat S_+ = a^\dagger\sqrt{N-\hat a^\dagger \hat a},\,\,\,\, 
  \hat S_- = \sqrt{N-\hat a^\dagger \hat a} \hat a,\\
  \hat S_x = \frac{1}{2}\left(\hat S_+ +\hat S_- \right),\,\,\,\,
  \hat S_z = \hat a^\dagger \hat a - \frac{N}2.
  \end{split}
\end{align}
For 
\begin{align}\label{eq:LMGapproximationcondition}
\langle \hat a^\dagger \hat a\rangle/N \ll 1 
\end{align}
the square roots can be expanded in the Taylor series to give
\begin{align}
  \hat S_x = \frac{\sqrt{N}}{2}\left(\hat a^\dagger +\hat a \right).
\end{align}
Plugging the above approximated operators into equation~\eqref{eq:LMG} and omitting the terms proportional to identity operator, we obtain
\begin{align}
  \hat H_\mathrm{LMG} = \omega \hat a^\dagger \hat a - \frac{g}{4}(\hat a^\dagger +\hat a)^2,
\end{align}
which describes a harmonic oscillator
\begin{align}
   \hat H_\mathrm{LMG} = \frac{\omega}{2}\hat P^2 +\frac{\omega}{2}\left(1- \frac{g}{\omega}\right)\hat X^2,
\end{align}
with a frequency (equivalently energy as $\hbar=1$) $\omega\sqrt{1-g/\omega}$ which becomes imaginary once $g>\omega$. If $g>0$, the above Hamiltonian is equivalent to Hamiltonian~\eqref{eq:qrm_col_down}; and if $g<0$, it corresponds to Hamiltonian~\eqref{eq:qrm_col_up}. According to the results from the previous section, we can expect that by going away from the critical point, it should be possible to create a rotated critical state exploiting the speed-up protocol.

Before we continue, let us closely examine the condition from equation~\eqref{eq:LMGapproximationcondition}. It naively seems that in the thermodynamic limit defined as $N\rightarrow \infty$ this condition always holds. However, in the previous section we have seen that close to the critical point $\langle \hat a^\dagger \hat a \rangle \rightarrow \infty$. Therefore, close to the critical point we cannot use the approximation~\cite{2013_HP_limit,2015LMG_shorcut}. Let us have a look however at its limitations. Using the approximation, the number of excitations $\langle \hat a^\dagger \hat a \rangle$ can be easily calculated. For $g>0$ (slowing-down) it becomes 
\begin{align}
  \langle \hat a^\dagger \hat a \rangle = \frac{1}{4\sqrt{1-\frac{g}{\omega}}},
\end{align}
which suggests that the approximation can be applied whenever
\begin{align}\label{eq:LMG_cond}
   \frac{1}{4}\frac{1}{\sqrt{1-\frac{g}{\omega}}} \ll N\,\, \rightarrow \,\, \frac{g}{\omega}\ll\frac{16 N^2-1}{16N^2}.
\end{align}
For $g<0$ (speeding-up) it becomes
\begin{align}
   \langle \hat a^\dagger \hat a \rangle = \frac{1}{4}\sqrt{1+\frac{|g|}{\omega}},
\end{align}
which suggests that the approximation can be applied whenever
\begin{align}\label{eq:LMG_su_cond}
  \frac{1}{4} \sqrt{1+\frac{|g|}{\omega}} \ll N \,\, \rightarrow \,\, \frac{|g|}{\omega} \ll 16N^2-1.
\end{align}

\begin{figure*}[htb!]
 \centering
\includegraphics[width=0.8\textwidth]{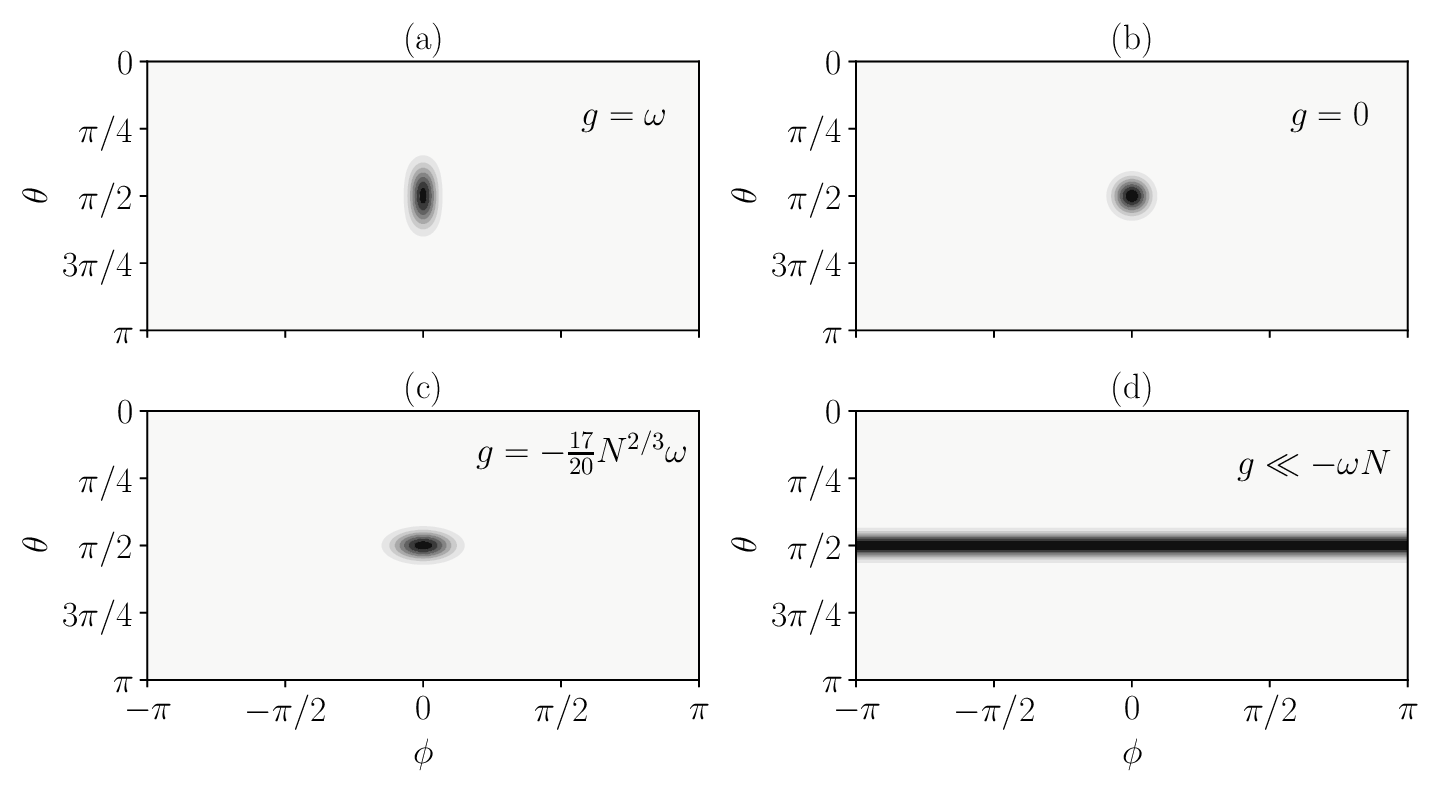}
\caption{Mercator projections of ground state SU(2) Husimi functions~\cite{1940Husimi} of the Lipkin-Meshkov-Glick model. (a), (b), (c), and (d), depict the critical ground state, the ground state for $g=0$, the rotated critical ground state for $g=-\frac{17}{20}N^{2/3}\omega$, and the maximally squeezed state, respectively. Here we have set $N=100$. The parametrization of the generalized Bloch sphere is such that $\theta = \pi/2$ and $\phi = 0$ corresponds to the collective spin-down state, and vertical and horizontal axes are $x$ and $y$ axes, respectively. See figure~\ref{fig:fig1} for comparison.}
\label{fig:fig4}
\end{figure*}

In order to determine the limits of applicability of the approximation in the speeding-up case, let us now take a step back and consider the Lipkin-Meshkov-Glick Hamiltonian~\eqref{eq:LMG}. In this case, near the critical point, we expect the state to be squeezed in $\hat S_y$ direction and anti-squeezed in $\hat S_x$ direction. Using the speed-up protocol, we will prepare a state squeezed in $\hat S_x$ direction and anti-squeezed in $\hat S_y$ direction. This can be seen in figure~\ref{fig:fig4}.

\begin{figure}[htb!]
 \centering
\includegraphics[width=0.4\textwidth]{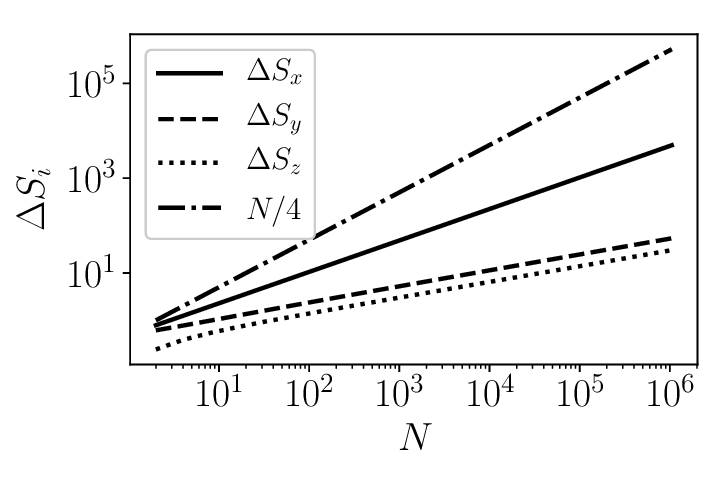}
\caption{Spin (anti) squeezing at the critical point of the Lipkin-Meshkov-Glick model as a function of $N$ for various components of the spin. The purple dashed line represents the maximally squeezed state.}
\label{fig:fig6}
\end{figure}

First of all, let us calculate the amount of (anti) squeezing in the critical state as a function of $N$. The results of the numerical simulations for $\Delta S_i$ as a function of $N$ for $g = 0.\overline{9}\omega$ is presented in figure~\ref{fig:fig6} (the bar notation indicates the recurring decimal). As we can see, the spin is indeed squeezed as expected but not maximally contrary to the approximation to the Holstein-Primakoff transformation. It can be shown that at the critical point (up to the leading order in $N$) $\Delta S_x = N^{2/3}/2$ so less than linearly with $N$ (see references~\cite{2004LMGfinitesizeexponents,2005LMGfinite,2009LMGFishersqueezing} for a rigorous derivation). 

\begin{figure*}[htb!]
 \centering
\includegraphics[width=0.8\textwidth]{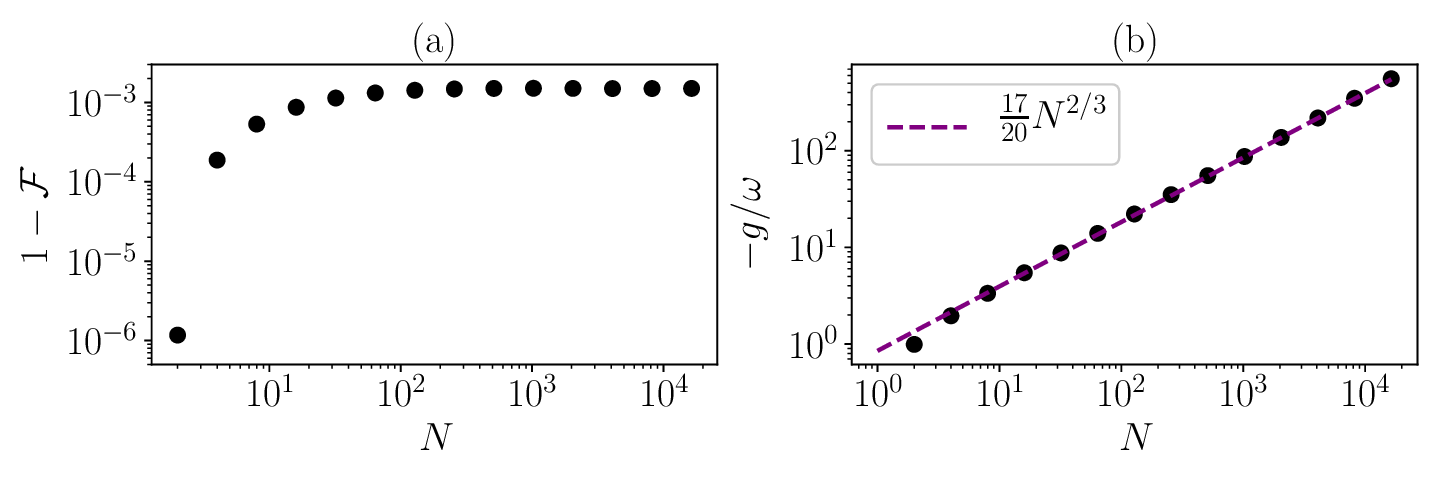}
\caption{Generating critical states by speed-up protocol. (a) shows infidelity, $1-\mathcal{F}$, as a function of $N$, and (b) shows relation between coupling strength and $\omega$ required for the protocol to work.}
\label{fig:fig7}
\end{figure*}

Once we know the critical state at $g=0.\overline{9}\omega$, we can find the corresponding $g$ which will lead to an equally squeezed state but rotated by $\pi/2$. Unfortunately, we do not have an analytical form of the wavefunction as in the case of the quantum Rabi model, and we have to numerically find the conditions. We do this by finding the ground state as a function of $N$ and $g/\omega$, rotating it by $\pi/2$ around the $z$ axis, calculating the overlap (fidelity) with the critical ground state, and finding its maximum. The results of the numerical simulations (exact diagonalization) are presented in figure~\ref{fig:fig7}.

First, we see that by increasing $N$ the fidelity defined as
\begin{align}
  \mathcal{F} = |\langle\mathrm{GS}(g_\xi)|\exp\left(i \frac{\pi}{2}\hat Sx \right)|\mathrm{GS}(g_\zeta)\rangle|^2,
\end{align}
where $|\mathrm{GS}(g_\xi)\rangle$ and $|\mathrm{GS}(g_\zeta)\rangle$ are the ground states with the same amount of squeezing but rotated by $\pi/2$ on the generalized Bloch sphere, saturates around $\mathcal {F}\approx 1-10^{-3}$. Although, we perform calculations for maximally $N=2^{14}$, it seems very unlikely that the fidelity will approach 1 once we further increase $N$. Second, similarly as for the quantum Rabi model, creating squeezing among larger number of particles requires higher interactions strengths [see figure~\ref{fig:fig7}(b)]. Note that by eliminating $N$ from the Lipkin-Meshkov-Glick model---i.e., making the model extensive---the interactions strength actually decreases. The numerical fit reveals that in order to prepare a rotated ground state one has to satisfy
\begin{align}\label{eq:LMG_su_cond2}
\frac{g}{\omega} = - \frac{17}{20}N^{2/3}.
\end{align}
The former together will the inability of creating a perfectly squeezed state is a confirmation of invalidity of the Holstein-Primakoff approximation close to the critical point. However, plugging the condition from equation~\eqref{eq:LMG_su_cond2} to equation~\eqref{eq:LMG_su_cond} yields
\begin{align}
   \frac{1}{4} \sqrt{1+\frac{17}{20}N^{2/3}} \ll N,
\end{align}
which means that by increasing $N$ we can use the approximation for the speed-up protocol. 

What is more, we can now use the condition from equation~\eqref{eq:LMG_su_cond2}, to find the approximate ground state at the critical point. We know that at the critical point the number of excitations is
\begin{align}
  \langle \hat a^\dagger \hat a\rangle = \frac{1}{4\sqrt{1-\frac{g}{\omega}}} \approx \frac{1}{4}\sqrt{1+\frac{17}{20}N^{2/3}}.
\end{align}
A simple calculation yields
\begin{align}\label{eq:propercondLMG}
  g = \frac{17 N^{2/3} }{17 N^{2/3}+20}\omega,
\end{align}
for which (the seemingly irrelevant 20 is very important here)
\begin{align}
  \langle \hat a^\dagger \hat a\rangle \approx 0.23 N^{1/3} \ll N.
\end{align}
Moreover, we can now calculate the (anti) squeezing
\begin{align}
\begin{split}
  \Delta^2 S_x &= \frac{N}{4}\langle \Xi|(\hat a^\dagger + \hat a)^2|\Xi \rangle = \frac{N}{4}\exp(2 \Xi) \\ &=  \frac{N}4 \frac{1}{\sqrt{1-\frac{g}{\omega}}} \le \frac{N^2}{4},
  \end{split}
\end{align}
where $|\Xi\rangle = \exp\{(\Xi/2)\hat a^{\dagger2} - (\Xi^*/2)\hat a^2\}|0\rangle$ with $\Xi = -\frac{1}{4}\log\{1-g/\omega\}$. Plugging in the condition from equation~\eqref{eq:propercondLMG} yields
\begin{align}\label{eq:newLMGcond}
  \Delta^2 S_x = \frac{1}{8} \sqrt{\frac{17 N^{8/3}}{5}+4 N^2} < \frac{N^2}{4}.
\end{align}
Now we are in the position to check whether this result predicts the correct amount of (anti) squeezing at the critical point. Up to the leading order with $N$, $\Delta S_x = N^{2/3}/2$ at the critical point. For large $N$, the expression from equation~\eqref{eq:newLMGcond} becomes,
\begin{align}
  \Delta S_x \approx 0.48 N^{0.6\overline{6}}= 0.48 N^{2/3}.
\end{align}
which is in a very good agreement (see also results from references~\cite{2004LMGfinitesizeexponents,2005LMGfinite,2009LMGFishersqueezing}). The above discussion indicates that the approximation to the Holstein Primakoff transformation can be applied to the Lipkin-Meshkov-Glick model under additional constraints, and the speed-up protocol can also be used. However, the fidelity of creating the critical state in this way seems to be limited to $\mathcal{F} \approx 1-10^{-3}$. 

So far, we have shown that in the case of the quantum Rabi model and Lipkin-Meshkov-Glick model it is possible to create a rotated critical state by going away from the critical point. With these results at hand we can proceed to finding the ramp that satisfies the adiabatic conditions and calculate the speed up.


\section{Adiabatic Quench}
In this section, we derive the form of the ramp that satisfies the adiabatic condition (see appendix~\ref{app:adcond}) for the quantum Rabi model and the Lipkin-Meshkov-Glick model.


\subsection{Quantum Rabi model}\label{sec:QRMadiabatic}
For the quantum Rabi model in the classical oscillator limit, $\frac{g}{g_c}\frac{{\omega}}{{\Omega}}\ll1$, the spin-up sector is separated from the spin-down sector and the transition of spin is forbidden. This can be explicitly seen by calculating $\dot{\hat H}$. A straightforward calculation yields
\begin{align}
  \dot{\hat H} = \frac{1}{2\Omega} \dot{g}(t)g(t)\left(\hat a +\hat a^\dagger\right)^2 \hat \sigma_z.
\end{align}
The above term does not couple spin-up and the spin-down sector since it only depends on $\hat \sigma_z$. This argument is valid, however, only for $g<g_c$. In the regime of $g>g_c$ the spin-down sector gets modified (double minimum phase), however, the symmetry of the wavefunction still does not allow for a transition from the spin-up to the spin-down sector once $\frac{g}{g_c}\frac{{\omega}}{{\Omega}}\ll1$ is satisfied. The ground state in the superradiant phase is a combination of $|\alpha\rangle\otimes |+\rangle$ and $|-\alpha \rangle \otimes |-\rangle$, where $|\beta\rangle$ is a coherent state of the harmonic oscillator and $|\pm\rangle$ is the eigen state of $\hat \sigma_x$. The symmetric superposition of these states gives a spin that points up, and an asymmetric superposition of these states gives a zero overlap with the symmetric squeezed state. Hence, there is no coupling between the sectors.
 
Having this result at hand, we can proceed to the calculation of the quench that satisfies the adiabatic condition [see equation~\eqref{eq:adiabaticcondition} in the Appendix]. Assuming $k=0$ for the ground state of the harmonic oscillator, a straightforward calculation yields
 \begin{align}
   \frac{\langle \psi_k|\dot{\hat H}|\psi_n\rangle}{E_n-E_k} =\mp \frac{1}{2 g_c^2} \frac{\dot{g}(t)g(t)}{1\mp \frac{g^2(t)}{g_c^2}}\ll \omega\sqrt{1\mp\frac{g(t)^2}{g_c^2}},
 \end{align}
 where the $\mp$ sign stands for the spin-down and the spin-up sector. Now, we will want to find a $g(t)$ that satisfies
 \begin{align}
   \dot{g}(t) = \frac{{2 \gamma \omega g_c^2}}{g(t)} {\left(1\mp \frac{g^2(t)}{g_c^2}\right)^{3/2}},
 \end{align}
 where $\gamma \ll 1$ (in agreement with reference~\cite{2020_CQM_Paris} but derived rigorously). A straightforward calculation yields
 \begin{align}
   g(t) = \frac{2 g_c \sqrt{\gamma t \omega (\gamma t \omega +1)}}{2 \gamma t \omega +1} 
\end{align}
 for the spin-down case, and
 \begin{align}
   g(t) =\frac{2 {g_c}\sqrt{\gamma  t \omega (1-\gamma t \omega)}}{ 1-2 \gamma t \omega }
 \end{align}
 for the spin-up case which becomes infinite for $t=1/2\omega\gamma$. The total time of the protocols can be calculated to be
 \begin{align}
   T_\xi = \int_0^{g_\xi} \frac{\mathrm{d}g}{\dot{g}(t)}  = \frac{1}{2 \gamma \omega \sqrt{1-\frac{{g}_\xi^2}{{g}_c^2}}}
\end{align}
for the spin-down case (again in agreement with reference~\cite{2020_CQM_Paris}) and
\begin{align}
  T_\zeta = \int_0^{g_\zeta} \frac{\mathrm{d}g}{\dot{g}(t)} = \frac{\sqrt{1+\frac{{g}^2_\zeta}{{g}_c^2}}-1}{2 \gamma \omega \sqrt{1+\frac{{g}_\zeta^2}{{g}_c^2}}}<\frac{1}{2\gamma \omega}
\end{align}
for the spin-up case. By using the condition from equation~\eqref{eq:speedupcond}, we can calculate the speed up
\begin{align}
  \frac{T_\xi}{T_\zeta} = \sqrt{\frac{1}{1-\frac{g^2}{{g}_c^2}}},
\end{align}
which is always greater than 1.


\subsection{Lipkin-Meshkov-Glick model}
In the case of the Lipkin-Meshkov-Glick model, adapting the calculations from the case of the quantum Rabi model we immediately arrive at
 \begin{align}
   \dot{g}(t) = \frac{ \gamma \omega^2}{4} {\left(1+ \frac{g(t)}{\omega}\right)^{3/2}}.
 \end{align}
 If $g(t)$ approaches the critical point, the solution is given by
 \begin{align}
 g(t) = \frac{\gamma t \omega^2\left(4+ \gamma t\omega\right)}{(\gamma t \omega +2)^2},
 \end{align}
whereas if $g(t)$ departs from the critical point, the solution is
\begin{align}
  g(t) = \frac{\gamma t \omega ^2 (4-\gamma t \omega )}{(\gamma t \omega -2)^2}.
\end{align}
The total time of the protocols can be calculated to be ($\Xi$ indicates the slow-down protocol)
\begin{align}\label{eq:LMGtottime}
\begin{split}
  T_\Xi &= \int_0^{g_\Xi} \frac{\mathrm{d}g}{\dot g (t)} = \frac{2}{\gamma \omega }{ \left(\frac{1}{\sqrt{1-\frac{g_\Xi}{\omega }}}-1\right)}\\& \stackrel{g_\Xi\sim \omega}{\approx} \frac{2}{\gamma \omega }{ \left(\frac{1}{\sqrt{1-\frac{g_\Xi}{\omega }}}\right)}
  \end{split}
\end{align}
for the ramping towards the critical point, and ($\Lambda$ indicates the speed-up protocol)
\begin{align}
  T_\Lambda =\int_0^{g_\Lambda} \frac{\mathrm{d}g}{\dot g (t)} = \frac{2}{\gamma \omega }{\left(1-\frac{1}{\sqrt{\frac{g_\Lambda+\omega }{\omega }}}\right)}<\frac{2}{\gamma \omega}
\end{align}
for the ramping away from the critical point. For the Lipkin-Meshkov-Glick model, the condition for preparing equally squeezed state becomes
\begin{align}
  \frac{g_\Lambda}{\omega} = \frac{g_\Xi}{g_\Xi-\omega},
\end{align}
and the speed-up turns out to be
\begin{align}
  \frac{T_\Xi}{T_\Lambda} = \frac{1}{\sqrt{1-\frac{g_\Xi}{\omega }}},
\end{align}
which is always greater than 1.

These results prove that by adiabatically going away from the critical point it is possible to build up the same amount of correlations in a shorter time. In both cases, driving the system away from the critical point will generate a rotated critical ground state. In order to make the true critical ground state out of it, one would have to set $g=0$ and wait for a $\pi/2$ rotation in the phase space which will last $\omega t = \pi/2$. 

\section{Applications in Quantum Metrology}
In this section, we present an application of the critical speeding-up protocol in quantum metrology and compare its performance with critical metrology. The starting point is a brief review of quantum metrology and critical metrology. Subsequently, we focus on the quantum Rabi model, and finally, we look at the Lipkin-Meshkov-Glick model.


\subsection{Quantum Metrology}
Quantum Metrology is a modern branch of physics that focuses on utilizing resources provided by quantum mechanics---such as nonclassical correlations---to overcome the standard quantum limit of precision~\cite{2009pezzesmerzient,2018pezzeRMP}. This limitation arises from the fact that if a system of $N$ uncorrelated particles is used in a measurement protocol, it is as if one was performing $N$ independent measurements on a single particle system. In this case, the optimal measurement sensitivity of Hamiltonian parameter estimation, here $\omega$, will be $\Delta \omega = 1/\sqrt{N}T$, where $T$ is the time of the process which imprints the information about the unknown parameter on the initial state. By correlating the particles in the system, it is possible to substantially increase the sensitivity and reach the ultimate Heisenberg limit. In the two-level systems [SU(2) systems], such as two-mode Bose-Einstein condensates, this limit becomes $\Delta \omega = 1/NT$, and in single-mode systems, such as single-mode radiation field, this limit becomes $\Delta \omega = \sqrt{1/8(\langle n\rangle + \langle n \rangle^2)}T$, where $\langle n \rangle$ is the average number of photons. In order to reach these quantum-enhanced bounds, one has to first prepare a suitable initial state. However, as this might be complicated in certain cases, it is possible to build the correlations simultaneously with imprinting the information about the unknown parameter. One such example which attracted much attention in recent years is the critical quantum metrology.

Critical metrology relies on driving the system in the vicinity of the critical point of a phase transition. Assuming that driving is adiabatic, the final (critical) ground state will depend on the unknown parameter.
This approach is motivated by the fact that critical states exhibit a high level of non-classicality (squeezing or spin-squeezing). Although these protocols can, in principle, exhibit Heisenberg or even super-Heisenberg scaling---this is a sensitivity that scales quadratically or even higher than quadratically with the number of particles or time---they will typically operate above the Heisenberg limit rendering them in fact in most of the cases impractical (it is very important to distinguish between the Heisenberg limit, which is a number deriving from the Heisenberg uncertainty principle, and the Heisenberg scaling, which is a quadratic scaling in time and the number of particles). This happens because reaching the Heisenberg limit requires having the optimal state from the beginning of the protocol. As the starting state of the critical metrology protocols is uncorrelated, critical metrology cannot reach the Heisenberg limit~\cite{2021_noshortcuttoCQM}. What is more, such protocols might last for a very long time since they are focused on approaching the critical points where the gap closes. 

Note that we do not use critical systems to first create a non-classical state and use it subsequently to imprint the information about an unknown parameter of some other process. The unknown parameter is a part of the Hamiltonian that we use to create the critical state. In this sense, we create correlations simultaneously with imprinting the information about the unknown parameter.


\subsection{Quantum Fisher Information}
An essential tool in quantum metrology is the quantum Fisher information as it is related to the sensitivity of a measurement of an unknown parameter, which we assume to be $\omega$, through the Cram\'er-Rao bound
\begin{align}
  \Delta \omega \geq \frac{1}{\sqrt{\mathcal{I_\omega}}},
\end{align}
where $\mathcal{I}$ is the quantum Fisher information defined as
\begin{align}
  \mathcal I_\omega = 4\left(\langle\partial_\omega \psi |\partial_\omega \psi\rangle - \langle \partial_\omega \psi| \psi\rangle^2 \right),
\end{align}
with $\partial_\omega \equiv \partial/\partial \omega$. If the Hamiltonian is composed solely of the term that imprints the information about the unknown parameter, for example $\hat a^\dagger \hat a$ in single mode systems or $\hat S_z$ in two mode systems, finding the optimal state and maximum value of the quantum Fisher information is a straightforward exercise. In the case of the single mode field, the quantum Fisher information becomes
\begin{align}
  \mathcal I_\omega = 4 T^2 \Delta^2 \hat a^\dagger \hat a,
\end{align}
which putting a restriction on the average number of excitations $\langle \hat a^\dagger \hat a\rangle \equiv \langle \hat n \rangle$ is maximized for a squeezed vacuum state and gives~\cite{2006Optimalgaussian}
\begin{align}
  \mathcal I_\omega = 8T^2(\langle \hat n \rangle^2+\langle \hat n \rangle).
\end{align}

In the case of a two-mode system, the quantum Fisher information becomes
\begin{align}\label{eq:u1HL}
  \mathcal I_\omega = 4 T^2 \Delta ^2\hat S_z,
\end{align}
which is maximized for a maximally entangled Greenberger–Horne–Zeilinger state and gives
\begin{align}\label{eq:SU2HL}
  \mathcal I_\omega = T^2 N^2.
\end{align}
The above equations for $\mathcal I_\omega$ set out the Heisenberg limit for single and two-mode systems. Most importantly, these two quantum Fisher informations can be increased by either increasing the time $T$ or increasing the number of particles or excitations.

For a more generic Hamiltonian
\begin{align}
  \hat H = \omega \hat H_\omega + \hat H_t(t),
\end{align}
where $\hat H_t(t)$ represents a general unknown-parameter-independent term of the Hamiltonian while $\omega\hat H_\omega$ is the term imprinting the information about $\omega$ ($\hat H_\omega$ itself does not depend on $\omega$), it can be shown that the quantum Fisher information is upper bounded by~\cite{2007PRL_GeneralizedLimits} 
\begin{align} \label{eq:QFIlimit}
\begin{split}
   \mathcal{I}_\omega  & \equiv 4\left(\langle \psi|\hat h^2|\psi\rangle - \langle\psi|\hat h|\psi\rangle^2 \right) \\ &\leq 4T^2\max_{|\phi\rangle}\left(\langle \phi|\hat H_\omega^2|\phi\rangle - \langle\phi|\hat H_\omega|\phi\rangle^2 \right).
   \end{split}
\end{align}
In the above expression, $|\psi\rangle$ is the initial state that does not depend on the unknown parameter, and $\hat h = i\hat U^\dagger\partial_\omega\hat U $ with $\hat U$ being the time evolution operator. According to the above expression, critical metrology protocols cannot reach the Heisenberg limit~\cite{2021_noshortcuttoCQM} as the instantaneous eigen state $|\psi\rangle$ cannot be the same as the optimal state $|\phi\rangle$. This however should not be seen as a major drawback as the derivation of the Heisenberg limit assumes instantaneous generation of an optimal state. Rather, the Heisenberg limit should be treated as a benchmark for a real protocol that takes into account the preparation of the initial state. However, a drawback of critical metrology protocol is a very slow generation of correlations (and excitations) which is a result of the closing of the energy gap. Therefore, the metrological resources are not exploited optimally as we will further see in the next sections.


\subsection{Quantum Rabi Model}
We begin with calculating the quantum Fisher information for the classical oscillator limit of the quantum Rabi model. A straightforward calculation yields
\begin{equation}
  \mathcal{I_\omega} = \frac{1}{8 \omega^2 \left(1 - \frac{g ^2}{g_c^2} \right)^2} \frac{g ^4}{g_c^4}. 
\end{equation}
From this expression, we see that near the critical point the quantum Fisher information is extremely large. However, let us have a detailed look at the origin of this apparent critical enhancement. First, due to the squeezing, this quantum Fisher information should scale quadratically with the number of (instantaneous) excitations which can be easily calculated to be
\begin{align}
  \langle \hat n \rangle = \sinh^2(\xi) \approx \frac{1}{4 \sqrt{1-\frac{g_\xi^2}{{g_c}^2}}}.
\end{align}
Second, this quantum Fisher information should also scale quadratically with time which, close to the critical point, is given by 
\begin{align}
  T_\xi \approx \frac{1}{2 \gamma \omega \sqrt{1-\frac{{g}_\xi^2}{{g}_c^2}}}.
\end{align}
A straightforward calculation shows that near the critical point the quantum Fisher information becomes
\begin{align}
   \mathcal{I_\omega} \approx {8 \gamma^2 \langle \hat n\rangle^2 T_\xi^2 }, 
\end{align}
with $\gamma\ll1$ as also shown in reference~\cite{2020_CQM_Paris} [see equation~\eqref{eq:u1HL} for comparison with the Heisenberg limit]. While the above quantum Fisher information exhibits Heisenberg scaling, it can only overcome the standard quantum limit once $\langle n \rangle>1/\gamma^2$. Assuming $\gamma = 0.01$, this corresponds to $\langle \hat n \rangle = 10000$ photons that can be achieved for $g_\xi/g_c \approx 0.9999999997$ which would require an extraordinary control over the coupling parameter.

An equally straightforward calculation for the speed-up protocol yields
\begin{equation}
  \mathcal{I}_\omega = \frac{1}{8 \omega^2 \left(1 + \frac{g ^2}{g_c^2} \right)^2} \frac{g ^4}{g_c^4} \leq \frac{1}{8 \omega^2} 
\end{equation}
As expected, the quantum Fisher information is limited and cannot be further enhanced by creating a more squeezed state. This happens because the speed-up protocol focuses on the creation of squeezing and not imprinting the information about the unknown parameter.

We want now to find out which strategy is better. Simultaneous creation of correlations and imprinting the information about the unknown parameter experiencing slowing-down (critical quantum metrology); or creating first correlations and then imprinting the information about the unknown parameter exploiting critical speeding-up (regular quantum metrology). To this end, we assume that we have a full control over the Hamiltonian parameters but a limited time which we set to $T = 35.4/\gamma \omega$, for which $g_\xi/g_c = 0.9999$ and $\langle \hat n \rangle \approx 17.7$. In the critical metrology approach, the quantum Fisher information can be calculated to be (we set $\gamma =0.01$)
\begin{align}
  \mathcal{I}_\omega \approx \frac{3.1}{\omega^2}\times 10^6.
\end{align}
In the protocol using the speed-up effect (alternatively using the bang-bang protocol), during the same amount of time, we can prepare an arbitrary number of photons and use the rest of the time to imprint the information about the unknown parameter on such a prepared state. Assuming we prepare at least $\langle \hat n \rangle \approx 17.7$ photons ($g_\zeta/g_c \approx 70.7$), the quantum Fisher information is 
\begin{align}
   \mathcal{I}_\omega > \frac{3.5}{\omega^2}\times 10^{9},
\end{align}
which is three orders of magnitude higher than the quantum Fisher information in the critical metrology approach. Keep in mind that it can be easily increased further by creating more photons, hence the inequality sign. Such an improvement is possible because in the critical approach to quantum metrology the correlations are being prepared very slowly. Therefore, critical quantum metrology is a suboptimal metrological strategy. Knowing that the approach exploiting fast generation of the correlated state first and then imprinting the information about the unknown parameter is superior to creating the correlated state very slowly and imprinting the information about the unknown parameter simultaneously, we can proceed to the discussion of the Lipkin-Meshkov-Glick model.


\subsection{Lipkin-Meshkov-Glick Model}
\begin{figure}[htb!]
 \centering
\includegraphics[width=0.4\textwidth]{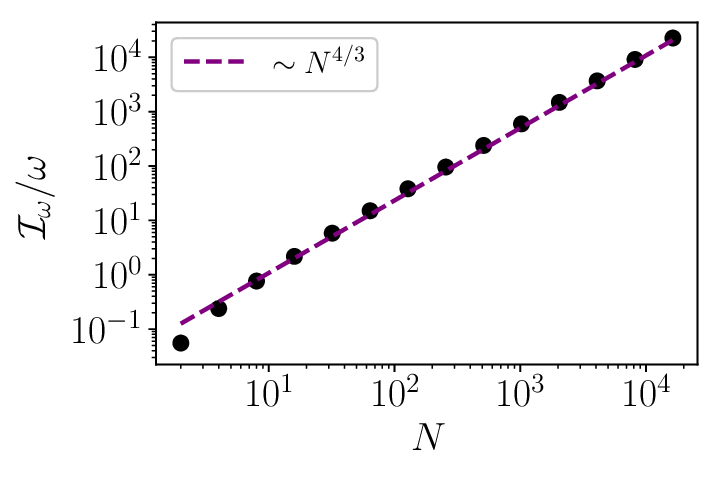}
\caption{Quantum Fisher information calculated at the critical point of the Lipkin-Meshkov-Glick model. Even though it exhibits greater than linear scaling with $N$ it is actually below the standard quantum limit. This happens because the critical state of the Lipkin-Meshkov-Glick model is only weakly squeezed as can be seen in figure~\ref{fig:fig4}(a) but the time of the protocol scales with $N$ which elevates the scaling.}
\label{fig:fig8}
\end{figure}

In the context of critical quantum metrology, the Lipkin-Meshkov-Glick is typically considered in the thermodynamic limit approximation of the Holstein-Primakoff transformation (see for instance reference \cite{2021_dynamicCQM}). However, the approximation is invalid close to the critical point as then the number of excitations $\langle \hat a^\dagger \hat a\rangle$ becomes infinite. Therefore it seems that the results derived for the critical quantum metrology in the Lipkin-Meshkov-Glick model might be incorrect. Since the squeezing of the critical state is not maximal as we showed in section~\ref{sec:LMG} (see figure~\ref{fig:fig6}) and scales only as $\Delta S_x \propto N^{2/3}$, critical quantum metrology with Lipkin-Meshkov-Glick model cannot reach the Heisenberg scaling (as it requires $\Delta S_i \propto N$). Also, using the adiabatic speed-up and creating the rotated (equally) squeezed state but much faster would not allow for the Heisenberg scaling. In order to see this, we begin now with the calculation of the quantum Fisher information. A calculation similar to that in the previous section shows
\begin{align}\label{eq:LMG_QFI_a}
  \mathcal{I}_\omega = \frac{1}{8 \omega^2 \left(1 - \frac{g }{\omega} \right)^2} \frac{g ^2}{\omega^2}.
\end{align}
On the other hand, the quantum Fisher information for the Lipkin-Meshkov-Glick model cannot be larger than~\cite{2007PRL_GeneralizedLimits}
\begin{align}\label{eq:LMG_QFI_r}
  \mathcal{I}_\omega = 4 \Delta^2 S_z T^2.
\end{align}
Therefore, we should expect to express the equation~\eqref{eq:LMG_QFI_a} in a similar form to equation~\eqref{eq:LMG_QFI_r}. In order to do this, we have to first calculate $\Delta^2 S_z$. An elementary calculations shows
\begin{align}\label{eq:LMGsz2}
\begin{split}
  \Delta^2 S_z &= \langle \Xi|\left(\hat a^\dagger \hat a -\frac{N}{2}\right)^2|\Xi\rangle -\langle \Xi|\left(\hat a^\dagger \hat a -\frac{N}{2}\right)|\Xi\rangle^2 \\&= \frac{g^2}{8 \omega ^2 \left(1-\frac{g}{\omega }\right)}
  \end{split}
\end{align}
By using expressions from equations~\eqref{eq:LMGtottime} and~\eqref{eq:LMGsz2}, we get
\begin{align}
  \mathcal{I}_\omega = \frac{\gamma^2}{4}\Delta^2 S_z T^2_\Xi,
\end{align}
as expected. Now at the critical point, it can be shown that the quantum Fisher information scales as
\begin{align}
  \mathcal{I}_\omega \propto \gamma^2 N^{2/3}T^2_\Xi,
\end{align}
which is in fact always below the standard quantum limit. Although a slight manipulation of this expression can lead to sub-Heisenberg scaling with the number of spins ($N^{4/3}$) by expressing the total time $T_\Xi$ as a function of $N$,
\begin{align}
  \mathcal{I}_\omega \propto \gamma^2 N^{2/3}T^2_\Xi \propto \frac{N^{4/3}}{\omega^2} \propto \gamma^4 \omega^2 T_\Xi^4,
\end{align}
it is still below the standard quantum limit [expressing $N$ as a function of $T_\Xi$ would lead to a super-Heisenberg scaling with time ($T_\Xi^4$)]. The results of the numerical simulations of the quantum Fisher information as a function of $N$ at the critical point are presented in figure~\ref{fig:fig8}.

On the other hand, using the speed-up protocol, one could easily prepare a very squeezed state in a short time and overcome the standard quantum limit. In an extreme case of preparing a maximally squeezed state [see figure~\ref{fig:fig4}(d)] it is possible to achieve the Heisenberg scaling of precision. However, from the practical point of view, the best strategy seems to be the very well-known quench beyond the critical point to create a spin-squeezed state~\cite{1993_Squeezed_Ueda} and then imprint the information about the unknown parameter.

\section{Conclusions and Outlook}
In this work we have presented an alternative protocol allowing for the preparation of critical (or squeezed) states that benefits from the opening of the energy gap and therefore is much faster than known adiabatic protocols. As the critical states are typically (anti) squeezed along certain phase space direction, the protocol relies on squeezing the initial state along an orthogonal direction in the phase space and (optional) rotation in the phase space by $\pi/2$. In this sense, the protocol drives the system away from the critical point which opens the energy gap and allows for the speed-up. Subsequently, we have derived the form of the adiabatic ramp that would allow for the speed-up in the quantum Rabi model and the Lipkin-Meshkov-Glick model, and show that the speed-up protocol is always superior to the adiabatic protocols relying on driving the system in the vicinity of the critical point. However, the limitation of the speed-up protocol might be the necessity to reach couplings way beyond the critical point. The presented protocol should be possible to apply in other physical systems exhibiting criticality, in particular, in the Dicke model whose two limits are the classical oscillator limit from equation~\eqref{eq:qrm_col}, and the fast oscillator limit equivalent to the Lipkin-Meshkov-Glick Hamiltonian~\eqref{eq:LMG}. However, we defer this topic for a future investigation. 

Next, we have applied the adiabatic speed-up protocol in quantum metrology and compared it with critical metrology protocols which attracted much attention in recent years. We showed that the adiabatic metrology exploiting the speed-up effect is superior to adiabatic critical quantum metrology. This can be easily explained as the critical quantum metrology relies on creating quantum resources (squeezing, spin-squeezing or entanglement) very slowly which means that it is a suboptimal metrological strategy. Even though critical quantum metrology can exhibit various scalings with $T$ and $N$ including Heisenberg and super-Heisenberg scaling~\cite{2018kubaPRX,2021garbecritical}, the ultimate precision will be always below the Heisenberg limit and it might be also below the standard quantum limit. After all, 
``it should be clear that scaling of the sensitivity is not \emph{per se} a desideratum. Any given instrument or measurement is judged by its sensitivity, not the scaling thereof''~\cite{2018qmreviewnoent}. In particular, we have shown that although critical quantum metrology with the Lipkin-Meshkov-Glick model can reach the super-Heisenberg scaling with $T$ (or greater than linear scaling with $N$), the absolute sensitivity is lower than the standard quantum limit. The optimal metrological strategies should create the quantum resources quickly, for example, by exploiting quenches into the critical phase to first prepare a suitable initial state and then to imprint the information about the unknown parameter or do it at the same time~\cite{2021exponentially}. As a matter of fact, quenches in the Lipkin-Meshkov-Glick model are a common way of preparing spin-squeezing~\cite{2011spinsqueezingreview,2012spinsqueezingmetrology} which can be later used in a metrological task~\cite{2014squeezingformetrology}.

In the context of quantum metrology, we discussed only the quantum Fisher information and did not calculate its classical version. However, as the states studied in this work are (Gaussian) squeezed vacuums, it can be easily shown that standard homodyne detection scheme or the measurements of spin components are the optimal measurements~\cite{2019Gaussianmetrologysqueezed} for which classical and quantum Fisher informations are equal.

The presented protocol could be used, in general, to create squeezed and spin-squeezed states and harness them subsequently for various other tasks. One of the most promising applications could be testing the performance of quantum heat engines exploiting the adiabatic speed-up protocol~\cite{2016_Fazio_criticalheatengine,2020Mossyengine}. 

\section*{Acknowledgements}
K.G. is pleased to acknowledge Lewis Ruks, Friederike Metz, and Thomas Busch for fruitful discussions. Simulations were performed using the open-source QuantumOptics.jl framework in Julia~\cite{KRAMER2018109}. This work was supported by the Okinawa Institute of Science and Technology Graduate University. K.G. acknowledges support from the Japanese Society for the Promotion of Science (P19792).

\appendix
\section{Adiabatic Condition}\label{app:adcond}
For the sake of completeness we derive the adiabatic condition. Following a textbook derivation, we consider a family of instantaneous eigen states
\begin{align}\label{eq:instanenous}
  \hat H(t)|\psi_n(t)\rangle = E_n(t)|\psi_n(t)\rangle.
\end{align}
Any given state can be decomposed in this basis according to
\begin{align}
  |\Psi(t)\rangle = \sum_n c_n(t)|\psi_n(t)\rangle.
\end{align}
Inserting this state into the Schr\"odinger equation yields (for the sake of brevity, we drop the explicit time dependence $f(t)\equiv f$)
\begin{align}
  i\sum_n\left(\dot{c}_n|\psi_n\rangle +c_n |\dot{\psi}_n\rangle\right) = \sum_n c_n E_n |\psi_n\rangle,
\end{align}
where the dot notation denotes a time derivative. Applying $\langle \psi_k|$ from the left side results in
\begin{align}
  i\dot{c}_k = E_kc_k-i\sum_n\langle \psi_k|\dot{\psi}_n\rangle c_n,
\end{align}
which can be rewritten factoring out the $k$th term from the sum in the following way\
\begin{align}\label{eq:inter}
  i\dot{c}_k = \left(E_k-i\langle \psi_k|\dot{\psi}_k\rangle\right)c_k-i\sum_{n\neq k}\langle \psi_k|\dot{\psi}_n\rangle c_n.
\end{align}
Taking the time derivative of equation~\eqref{eq:instanenous} yields
\begin{equation}
  \dot{\hat H}|\psi_n\rangle + \hat H|\dot{\psi}_n\rangle = \dot{E}_n|\psi_n\rangle + E_n|\dot{\psi}_n\rangle.
\end{equation}
Applying $\langle \psi_k|$ with $k\neq n$ from the left side gives
\begin{equation}
  \langle\psi_k|\dot{\hat H}|\psi_n\rangle +E_k\langle \psi_k|\dot{\psi}_n\rangle = {E}_n\langle\psi_k|\dot{\psi}_n\rangle,
\end{equation}
and after rearranging the terms
\begin{align}
  \langle \psi_k|\dot{\psi}_n\rangle = \frac{\langle \psi_k|\dot{\hat H}|\psi_n\rangle}{E_n-E_k}.
\end{align}
Plugging the above equation into~\eqref{eq:inter} yields
\begin{align}
   i\dot{c}_k = \left(E_k-i\langle \psi_k|\dot{\psi}_k\rangle\right)c_k-i\sum_{n\neq k}\frac{\langle \psi_k|\dot{\hat H}|\psi_n\rangle}{E_n-E_k} c_n.
\end{align}
It can be shown that in the quantum Rabi model (also in the Lipkin-Meshkov-Glick model) $\langle \psi_k|\dot{\psi}_k\rangle=0$, therefore if
\begin{align}\label{eq:adiabaticcondition}
  E_k \gg \left|i\sum_{n\neq k}\frac{\langle \psi_k|\dot{\hat H}|\psi_n\rangle}{E_n-E_k}\right|,
\end{align}
all the $|c_k|$ are constant throughout the entire process and hence the evolution is adiabatic.

\end{document}